\documentclass[twocolumn,preprint]{aastex62}
\usepackage[utf8x]{inputenc}
\usepackage{amsmath,amsfonts,amssymb,mathrsfs,bm,enumerate,graphicx}

\usepackage[T1]{fontenc}
\usepackage[all]{hypcap}
\usepackage{ulem}

\hypersetup{breaklinks,colorlinks,citecolor=blue,linkcolor=magenta}

\defcitealias{2019arXiv190802326D}{DLL20}

\newcommand{\pderiv}[1]{\frac{\partial}{\partial #1}}
\newcommand{\ppderiv}[2]{\frac{\partial #1}{\partial #2}}

\newcommand{\be}{\begin{eqnarray}}
\newcommand{\ee}{\end{eqnarray}}

\shorttitle{Disk Inner boundaries}
\shortauthors{{Dempsey},  {Mu\~{n}oz}, \& {Lithwick}}
\begin{document}
\title{Inner Boundary Condition in Quasi-Lagrangian Simulations of Accretion Disks}
\author[0000-0001-8291-2625]{Adam M. Dempsey}
\affiliation{Theoretical Division, Los Alamos National Laboratory, Los Alamos, NM 87545, USA}
\affiliation{Center for Interdisciplinary Exploration and Research in Astrophysics (CIERA)
and
Department of Physics and Astronomy
Northwestern University \\
2145 Sheridan Road
Evanston, IL 60208
USA}
\author[0000-0003-2186-234X]{Diego Mu\~{n}oz}
\affiliation{Center for Interdisciplinary Exploration and Research in Astrophysics (CIERA)
and
Department of Physics and Astronomy
Northwestern University \\
2145 Sheridan Road
Evanston, IL 60208
USA}
\author{{Yoram Lithwick}}
\affiliation{Center for Interdisciplinary Exploration and Research in Astrophysics (CIERA)
and
Department of Physics and Astronomy
Northwestern University \\
2145 Sheridan Road
Evanston, IL 60208
USA}

\email{adempsey@lanl.gov}

\begin{abstract}

In simulations of viscously evolving accretion disks, 
the inner boundary condition is particularly important.
If treated incorrectly, it  induces incorrect behavior very quickly, because
the viscous time is shortest near the inner boundary. 
Recent work has determined the correct inner boundary in Eulerian simulations.
But in  quasi-Lagrangian simulations  (e.g., SPH, moving mesh, and mesh-less), 
where the inner boundary is modeled
 by removing mass within a finite zone, the inner density  profile typically  becomes 
 anomalously depleted. 
Here we show how the boundary condition should be applied in such codes,  
via a simple modification 
of the usual approach: when one removes mass, one must speed up the remaining material so that the disk's angular momentum is unchanged. We show with both 1D and  2D moving-mesh ({\footnotesize AREPO}) simulations that this
scheme works as desired in viscously evolving disks. It produces no spurious density depletions
and is independent of the mass removal rate, provided that the disk is adequately resolved and that the mass removal rate is not so extreme as to trigger instabilities.
This ``torque-free'' mass removal technique permits the use of quasi-Lagrangian codes to simulate viscously evolving disks, while  including a variety of additional effects. As an example, 
we apply our scheme to a 2D
simulation of an
accretion disk perturbed by a very massive planet, in which the disk is evolved to viscous steady state.

\end{abstract}


\section{Introduction} \label{sec:intro}

In simulating  accretion disks, it is common to adopt an
inner boundary that lies far outside of the true disk inner edge. 
This allows
 the computationally expensive---and often uncertain---dynamics of the innermost
regions to be avoided. 
In this {\it Letter}, we work out the correct inner boundary condition, 
 such that a disk with a large inner boundary mimics one with a boundary further
 in. 
 We then show how it may be applied in a simple way to
   quasi-Lagrangian codes.
We focus
 on a viscously evolving disk that is perturbed by an embedded
planet, but discuss some other applications in \S \ref{sec:pdisk}.

Since the viscous time decreases inwards \citep{1974MNRAS.168..603L}, an incorrect
inner boundary condition can quickly lead to an incorrect density profile in the inner
parts of the simulation---long before the density in the vicinity of the planet can evolve viscously.\footnote{
 We do not address the {\it outer} boundary   because its effects are
 less pernicious: they are only important on timescales longer than the viscous
 time at the outer boundary.  See also \citet{2019arXiv190802326D}, who work out the outer
 boundary condition needed for such long timescales.
}
But that short inner viscous time also suggests the principle that should be applied to
determine the correct inner boundary: that the inner parts of the disk should be locally in
 steady-state, with a radially independent mass accretion rate.

In Eulerian simulations of accretion disks, inner boundary conditions 
may be applied at a fixed spatial location, as done in our previous work \citet[][hereafter \citetalias{2019arXiv190802326D}]{2019arXiv190802326D};
see also \S\ref{sec:1d}. 
But in quasi-Lagrangian simulations, it is more natural to remove mass
within an  inner zone, as has been done in smoothed particle hydrodynamics \citep[SPH; e.g.,][]{1995MNRAS.277..362B};
 in meshless codes \citep[e.g.,][]{2017arXiv171201294H}; and in
 structured and unstructured moving-mesh codes \citep[][]{2014ApJ...783..134F,2016ApJ...827...43M}.
In the aforementioned papers,  mass is removed without changing the velocity of the remaining material.\footnote{\citet{1995MNRAS.277..362B} also correct for the discontinuity in density that develops across
the accretion radius in SPH simulations due to a lack of particles at smaller radii; see \S\ref{sec:conclusions} for further details.}
This method, as shown below,  artificially exerts a torque on the disk, 
which leads to an incorrect inner density profile within the  computational domain.\footnote{\cite{2004ApJ...611..399K} apply mass removal in a correct way. But puzzlingly, their resulting inner density profile is not correct. We discuss their work  further  in \S\ref{sec:conclusions}.}

Artifacts induced by artificial mass removal have recently given rise to a controversy 
in the context of disks with embedded stellar binaries.
\cite{2017MNRAS.469.4258T} claimed that the inner mass removal procedure
can be crucial for determining the sign of the torque on the binary.
If that were true, the surprising  result that 
 accreting binaries {\it expand} \citep{2019ApJ...871...84M,2020ApJ...889..114M} 
 could be an artifact of incorrect mass removal. 
While follow-up studies \citep{2019ApJ...875...66M,Duffell:2019vu}
were unable to confirm
 the \citeauthor{2017MNRAS.469.4258T} claim, it is still desirable to devise mass removal
methods that introduce no artifacts at all. In this work, we describe and
implement such a method in the moving-mesh code {\footnotesize AREPO}
\citep{2010MNRAS.401..791S}.

\section{Torque-free Inner Boundary}

\subsection{Why the inner boundary should be torque-free}
\label{sec:why}
The 1D equations of mass and angular momentum conservation for a viscous accretion disk,
in the absence of a planet, are \citep{1974MNRAS.168..603L}
\be
2 \pi r \ppderiv{\Sigma}{t}&-&\ppderiv{\dot{M}}{r} = 0  \ ,\label{eq:sig} \\
2 \pi r \ppderiv{\Sigma \ell}{t}&+&\pderiv{r} \left( F_\nu - \dot{M} \ell \right) = 0 \ , \label{eq:mom} 
\ee
where $\ell = r^2 \Omega(r)$ is the specific angular momentum; $\Omega\propto r^{-3/2}$  is the
Keplerian angular speed i.e., ignoring the pressure correction; $\dot{M}(r)$ is the radial mass flux; and  
$F_\nu = 3 \pi \nu \Sigma \ell$ is the
viscous angular momentum flux, 
where $\nu$ is the kinematic viscosity.
One may solve the above equations  for 
\be  \label{eq:mdot}
\dot{M}= \left( \frac{d \ell}{dr} \right)^{-1} \frac{\partial F_\nu}{\partial r} \ ,
\ee 
and that expression  may then be inserted into equation (\ref{eq:sig}), yielding a
diffusive partial differential equation for $\Sigma$. 

The fact that the viscous time decreases inwards implies that, 
if $\Sigma$ at some fiducial radius  is varying due to viscous evolution, 
then the $\Sigma$  at much smaller radii 
   can be determined by setting
$\partial_t=0$ in the above equations. Doing so, we find
\be
\dot{M} &=& {\rm const. \ in\ }r  \ , \\
\Delta T&\equiv& F_\nu-\dot{M}\ell = {\rm const. \ in\ }r  \ . \label{eq:deltat_def}
\ee
  The quantity 
$\Delta T$ is the net torque, i.e., it is the sum of two angular momentum fluxes:
the (outwards) viscous flux and the (inwards) advective flux. 
As a result, the surface density far inside of the fiducial radius is given by
\be \label{eq:sigma}
\Sigma = \frac{\dot{M}}{3 \pi \nu } + \frac{\Delta T}{3 \pi \nu \ell}  \ . \label{eq:sigm} 
\ee
As is well-known, the $\Sigma$ profile in a steady-state disk is a sum of two terms, with
two arbitrary constants, $\dot{M}$ and $\Delta T$
 \citep{1974MNRAS.168..603L}.
 
  The second term in 
Equation~(\ref{eq:sigm}) falls with distance faster than the first.
Therefore, while these
two terms  are comparable at the disk's true inner edge
\citep{1974MNRAS.168..603L}, 
the second one becomes subdominant further out.
Consequently, in order to apply an inner boundary condition far beyond the true inner
edge, one should enforce 
$\Delta T=0$ at the boundary.  
We call this a  ``torque-free'' boundary. 

For future reference, we  refer to the  $\Sigma$ solution with $\Delta T=0$
 as the ZAM
 (``zero angular momentum flux'')
   solution, i.e., 
\be \label{eq:zam}
\Sigma_Z = \frac{\dot{M}}{3 \pi \nu} \ .
\ee

\subsection{Implementation of torque-free inner boundary} \label{sec:implementation}

A torque-free inner boundary is simple to implement when  solving the 1D
equations  
  with a  finite difference method. 
Substituting
 Equation (\ref{eq:mdot}) into the relation $F_\nu-\dot{M}\ell=0$ yields
a relationship between $F_\nu$ and $\partial_r F_\nu$, and hence between $\Sigma$ and $\partial_r \Sigma$, that can be applied at the inner
boundary (a Robin boundary condition). 
Essentially the same technique can also be implemented in 2D Eulerian hydrodynamical codes with little difficulty, e.g., with {\footnotesize FARGO3D} \citep{2016ApJS..223...11B}, as done in \citetalias{2019arXiv190802326D}.

On the other hand, in quasi-Lagrangian codes
it is more natural
to remove mass throughout an extended zone.
In most previous work, mass is removed  either by eliminating entire particles \citep[e.g.,][]{1995MNRAS.277..362B}, 
or by draining mesh cells, reducing their mass and 
momenta by the same fraction \citep[e.g.,][]{2014ApJ...783..134F,2016ApJ...827...43M}.
In both cases, 
the velocity of the remaining
fluid is left unchanged.
This technique, which we call 
 ``naive mass removal,''
 violates the torque-free inner boundary condition,
because when  momentum is removed, an angular momentum sink term must be added to 
 Equation~(\ref{eq:mom}). As a result, in viscous steady state the
value of $\Delta T$ at the outer edge of the mass removal zone (which we denote $r_0$) 
will be non-zero. Note that $r_0$ corresponds to what we have been calling the inner boundary, 
because it is the inner  boundary of the region where the physics is properly
modelled (i.e., no 
mass removal.)  Of course, the region inside of $r_0$ (the mass removal zone) is still modeled with the code.

One may enforce $\Delta T=0$ at $r_0$ by ensuring that there are no sources
or sinks of angular momentum in the mass removal zone, so that Equation~(\ref{eq:mom})  remains unchanged.
In quasi-Lagrangian codes, this can be accomplished by increasing the azimuthal velocity
of a mesh cell  every time its mass is reduced, such that its angular momentum is preserved.
We call this technique ``torque-free mass removal,'' and we shall investigate it in more detail
in the remainder of this {\it Letter}.

\begin{figure}
\centering
\includegraphics[trim={0.3cm 0.3cm 0.22cm 0},clip,width=.48\textwidth]{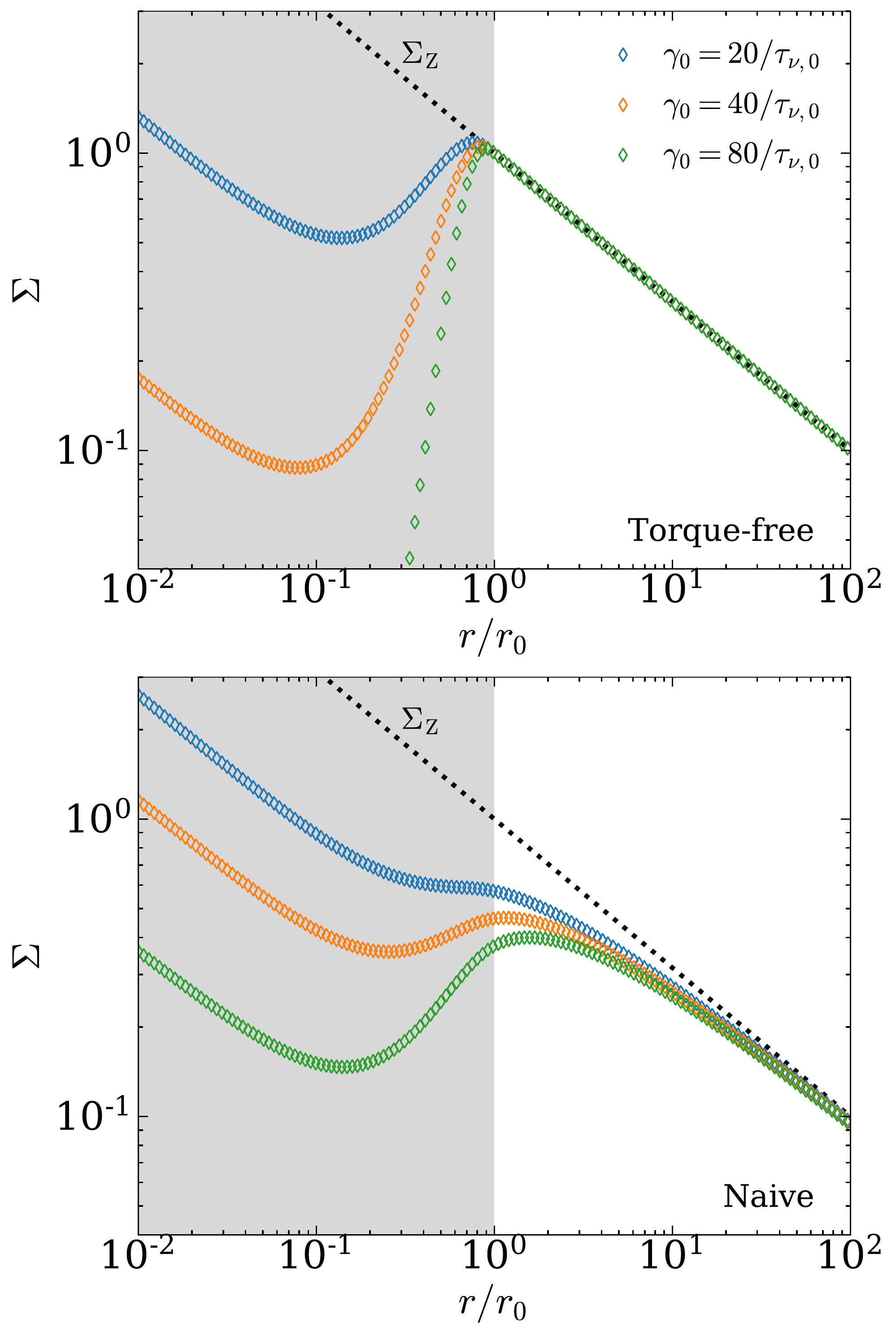}
\caption{
Steady-state profiles from 1D simulations of torque-free (top) and naive (bottom) mass removal.
 Three  values of $\gamma_0$ are shown, with values in the inset in units of $\tau_{\nu,0}$, which is the viscous 
 time at $r=r_0$.}
  \label{fig:1d_gamma_beta}
\end{figure}

\subsection{Torque-free versus naive mass removal  in 1D}\label{sec:1d}
It is instructive to demonstrate the difference between torque-free and naive mass removal in 1D.
Mass removal can be modeled in 1D simulations by modifying
Equation~(\ref{eq:sig}) to 

\begin{equation}
\label{eq:sigdotrem}
    2 \pi r \ppderiv{\Sigma}{t}  - \ppderiv{\dot{M}}{r} = - \gamma(r) 2 \pi r \Sigma , 
\end{equation}
where
$\gamma$ is a mass removal function that vanishes for $r\geq r_0$. 
For $r<r_0$, we choose the form
\be \label{eq:gamma}
\gamma(r) =  
\gamma_0(1 - r/r_0)^2 
\ee
which rises  to $\gamma_0$ at the origin; here,    $\gamma_0$ is a  constant with units of inverse time.

To model torque-free mass removal, we evolve Equations (\ref{eq:mom}) and (\ref{eq:sigdotrem})
with a finite difference code. 
We initialize with the ZAM value for $\Sigma$, set the outer boundary condition to have
fixed $\dot{M}$ (via Equation (\ref{eq:mdot}) with $\dot{M}$ specified), 
and run until the $\Sigma$ profile reaches steady state.
We also set $\nu={\rm const}\times r^{1/2}$ and adopt the aforementioned Robin inner boundary
condition at $r=0.01 r_0$.

 The result is shown in Figure~\ref{fig:1d_gamma_beta} (top panel),
for three values of $\gamma_0$.  We see that torque-free mass removal 
produces the correct (ZAM) profile at $r>r_0$, independent of $\gamma_0$, as desired.  
Inside of $r_0$, the depression in $\Sigma$ increases with increasing $\gamma_0$.
As discussed below, in 2D simulations one typically wishes to 
choose $\gamma_0 \sim 10$-$100/\tau_{\nu,0}$, where $\tau_{\nu,0}$
is the viscous time (i.e., $r^2/\nu$) at $r_0$.

The bottom panel of Figure \ref{fig:1d_gamma_beta} contrasts what happens with naive
mass removal.  
For that panel we proceed as before, except that we add  
 the sink term $-\gamma 2\pi r\Sigma \ell$ to Equation (\ref{eq:mom}), for reasons discussed above.
We infer from the plot that naive mass removal depresses the density profile outside of $r=r_0$,  relative to the correct $\Sigma_Z$. This depression worsens as $\gamma_0$ increases.
In general, deviations from $\Sigma_Z$  indicate that the inner edge of the disk is
being incorrectly torqued -- in this case due to the mass removal algorithm.

\section{Torque-Free Mass Removal in {\footnotesize AREPO}} \label{sec:2d}

We run 2D quasi-Lagrangian simulations with
  {\footnotesize AREPO} \citep{2010MNRAS.401..791S,2016MNRAS.455.1134P},
  which solves the Navier-Stokes equations  \citep{2013MNRAS.428..254M}
 on a moving mesh. The mesh 
 is constructed from a Voronoi tessellation of points that
 move with the fluid.
  Our setup is broadly similar to that of \citet{2014MNRAS.445.3475M}:
  the viscous stress tensor is implemented via a kinematic shear
 viscosity, parametrized with the  \citet{1973A&A....24..337S} $\alpha$ prescription,
 and the equation of state is locally isothermal with constant aspect ratio. Here,
 we use  $\alpha = H/r= 0.1$.
 The central potential is softened inside of $r_0$ with a spline function \citep{2001NewA....6...79S}
 such that outside of $r_0$ the potential is exactly Keplerian. 
See Appendix \ref{sec:app_2d} for further computational details, including
 a resolution study.

We implement torque-free mass removal as described in \S\ref{sec:implementation}.
At every timestep, we reduce the mass of a cell at the rate given by 
Equation (\ref{eq:gamma}), 
while preserving the cell's angular momentum, which has the effect of 
 boosting the cell's azimuthal velocity.
Our initial $\Sigma$ profile  has a  cavity out to $3 r_0$, and an outer exponential cutoff at $r = 100 r_0$. 

\begin{figure}
\centering
\includegraphics[trim={0.3cm 0.3cm 0.22cm 0},clip,width=.48\textwidth]{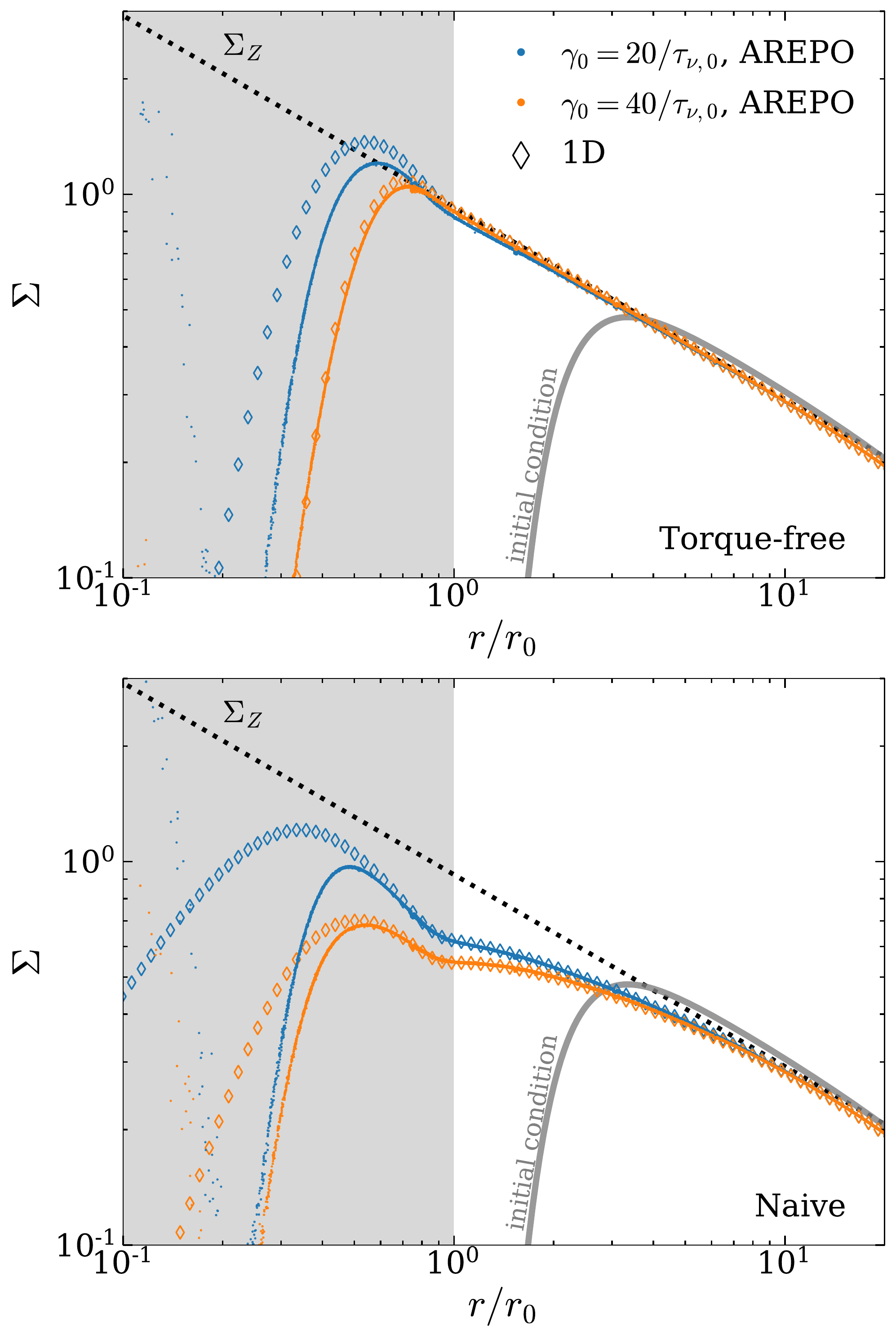}
\caption{
Surface densities resulting from torque-free mass removal (top) and naive mass removal (bottom).
We plot the density of each cell in {\footnotesize AREPO}  with a dot.
The fact that there is little scatter, except at very small $r$, indicates that the
$\Sigma$ profile is very axisymmetric. The 1D profiles here differ  slightly from those in Figure 
\ref{fig:1d_gamma_beta} within the mass removal zone because of the softened potential
used here.
}
      \label{fig:2d_1d_comp}
\end{figure}

In Figure \ref{fig:2d_1d_comp} (top panel), we compare
the {\footnotesize AREPO} 2D solutions to 
those obtained by solving the 1D equations, for 
 $\gamma_0=20/\tau_{\nu,0}$ and $=40/\tau_{\nu,0}$.
 We find excellent agreement between the two methods
at $r>r_0$. 
 The simulations in the figure have been run to a time equal to 800 orbits at
$r_0$, which corresponds to a viscous time at $r \approx 3 r_0$. But the $\Sigma$ 
profiles cease to evolve visibly at times $\gtrsim$ 400  orbits at $r_0$.
There is a modest discrepancy
 between {\footnotesize AREPO} and  1D results for the case with lower $\gamma_0$  due to the larger discretization errors at smaller $r$ (see also below).

The bottom panel shows a similar comparison, but for naive mass removal---both in {\footnotesize AREPO}
and in the 1D code. As before, the results of the two codes agree at $r>r_0$. 
For the simulations shown, naive mass removal gives a noticeable density depletion relative to the
ideal (ZAM) solution out to $r\sim 3$--5$r_0$. 
Similar depressions are often seen in SPH simulations of protoplanetary disks that implement naive mass removal via a sink particle
\citep[e.g.,][]{2018MNRAS.473.1603H,2018PASA...35...31P}, but are also apparent in simulations using other methods \citep[e.g.,][]{2004ApJ...611..399K,2017MNRAS.469.4258T}.

How should one choose $\gamma_0$? 
There are two competing considerations.  If $\gamma_0$ is too small, then too little mass is removed, 
 obviating the reason  for using mass removal in the first place. 
Conversely, if $\gamma_0$ is too large, we find that the disk develops
non-axisymmetries.  For example,  a simulation with $\gamma_0 \gtrsim 80/\tau_{\nu,0}$ but otherwise
the same as  Figure \ref{fig:2d_1d_comp} develops an $m=2$ pattern 
at $r_0$ (see Appendix \ref{sec:app_res}). 
We have not identified the origin of the instability, but note
that for large $\gamma_0$ the density gradient becomes large near $r_0$, 
and hence the disk can experience Rayleigh \citep{1961hhs..book.....C} or Rossby wave instabilities \citep{2000ApJ...533.1023L,1999ApJ...513..805L}.

\begin{figure*}
\centering
\includegraphics[trim={0.3cm 0.3cm 0.22cm 0},clip,width=.88\textwidth]{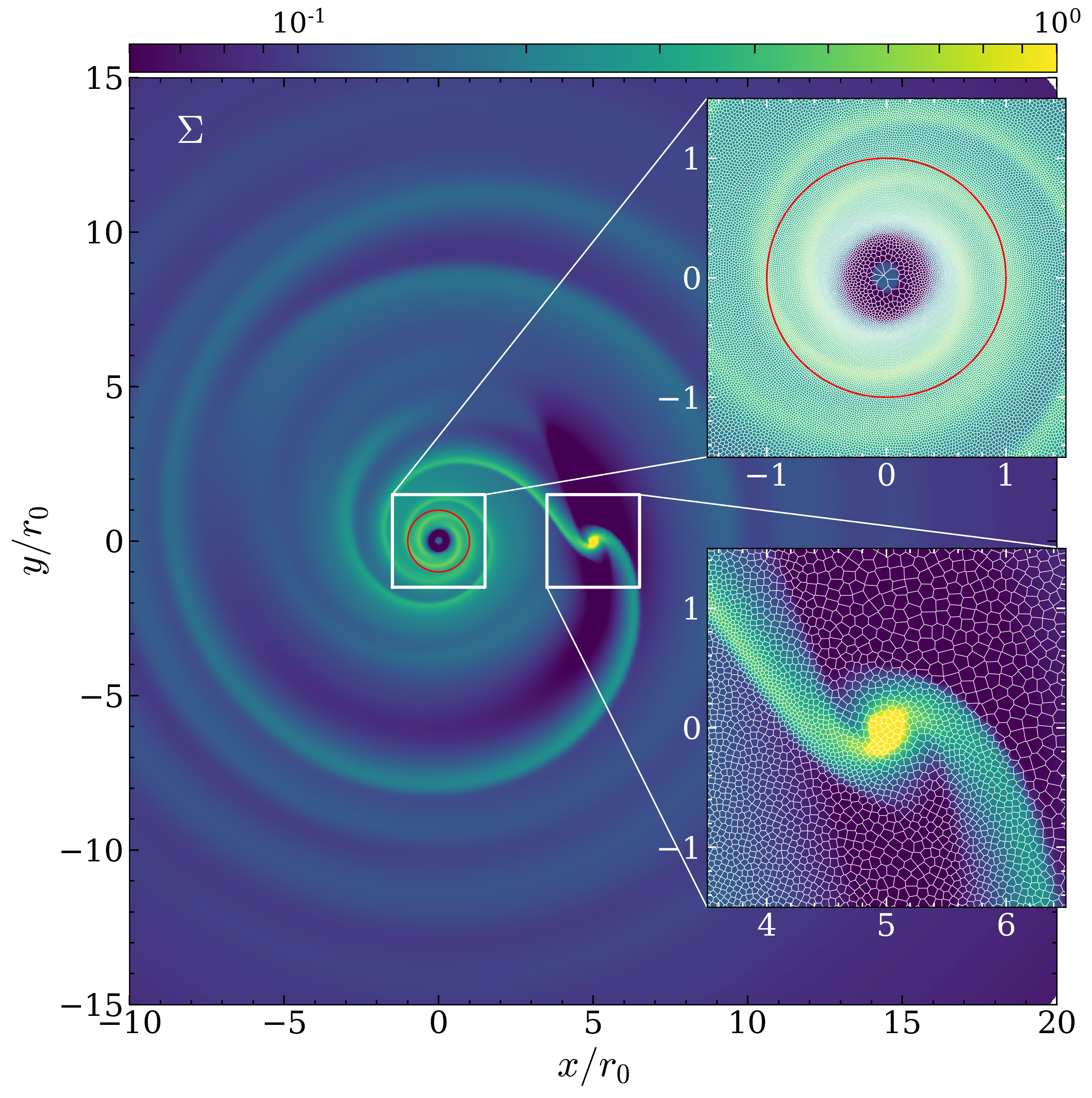}
\caption{Disk surface density for a planet with mass ratio $5 \times 10^{-3}$ after
the disk has reached viscous steady state. The red circle is at $r=r_0$. 
 The insets around the star and the planet show the Voronoi mesh, 
 which is refined with mass-refinement around the planet, and cellsize-refinement around the star (see Appendix \ref{sec:app_ref}).
  }
  \label{fig:planet_2d}
\end{figure*}

\section{Steady-state Disk perturbed by a Planet} \label{sec:pdisk}

We apply torque-free mass removal to the situation studied in \citetalias{2019arXiv190802326D}: 
a planet is placed on a fixed orbit within a viscous disk, 
and then the disk
is evolved long enough to reach viscous steady state beyond the planet's orbit.

We take the planet's mass to be $5 \times 10^{-3}$ times that of the star, and place it on a circular orbit  at $5 r_0$. We set the disk parameters to be the same as
in Figure \ref{fig:2d_1d_comp}, 
with $\gamma_0 = 40/\tau_{\nu,0}$, and work in the stellar-centric frame by including
an indirect potential. 
We  run the simulation with {\footnotesize AREPO}, using torque-free mass removal, to a time of
 $\sim 1,100$ planetary orbits, which is the
viscous time at $\sim 18 r_0$.

Figure \ref{fig:planet_2d} shows a snapshot of the resulting surface density and Voronoi mesh.
The planet excites  spiral density waves, which
 torque the gas, altering the $\Sigma$ profile (\citealt{1980ApJ...241..425G}; \citealt{1982Icar...52...14L}; \citealt{2017PASJ...69...97K}; \citetalias{2019arXiv190802326D}).
 Note that we do not employ wave-killing in the simulation \citep[e.g.,][]{2006MNRAS.370..529D}.
Despite that, the waves launched by the planet do not reflect off of the inner
 boundary, as evidenced by the absence of leading spiral arms in the figure.

\begin{figure}
\centering
\includegraphics[trim={0.3cm 0.3cm 0.22cm 0},clip,width=.42\textwidth]{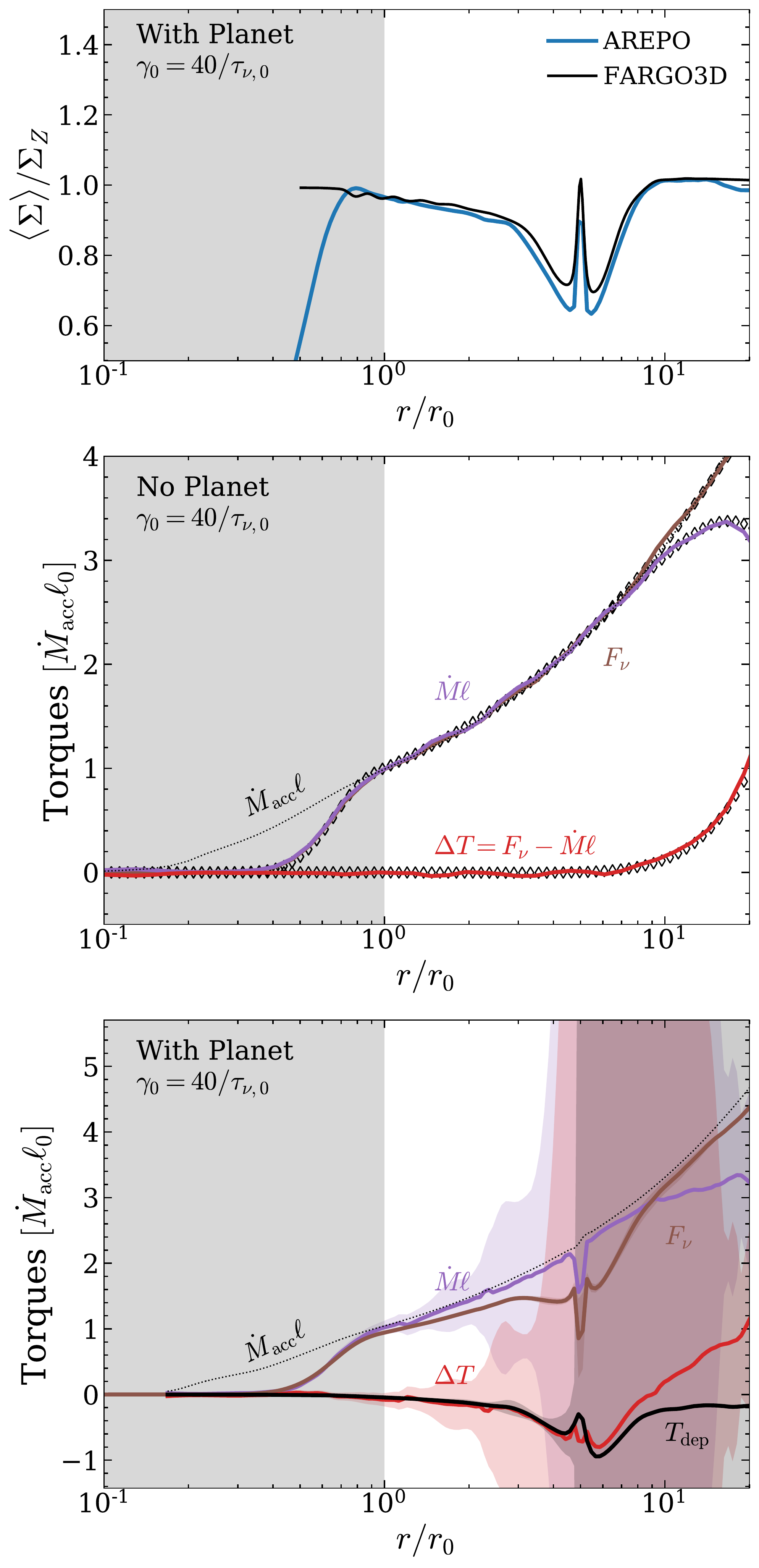}
\caption{
The top panel shows the azimuthally averaged $\Sigma$ profile for our planet runs with {\footnotesize AREPO} (blue line) and  {\footnotesize FARGO3D} (black line);
 $\Sigma$  is scaled to the ZAM profile (Equation \ref{eq:zam}), taken
at an $\dot{M}$ equal to the one measured in the simulations ($=\dot{M}_{\rm acc}\approx 0.35 \times 3\pi\nu_0\ell_0$). 
The middle  panel shows the torques in the  planet-less {\footnotesize AREPO} simulation from Figure \ref{fig:2d_1d_comp} (colored lines), as well as those from the 1D code (diamonds).
The bottom panel shows the torques in the {\footnotesize AREPO} simulation from the top panel, time-averaged over $50$ planet orbits. 
We describe how we compute the azimuthally averaged torques and $\Sigma$ profile in Appendix \ref{sec:fluxes}.
    \label{fig:planet_vss}
}
\end{figure}

 Figure \ref{fig:planet_vss} (top panel)  shows the $\Sigma$ profile of the disk.
The planet is massive enough to open a modest gap around its orbit.
Furthermore, $\Sigma$ remains below the ZAM profile  (Equation \ref{eq:zam}) nearly to $r_0$, due to the 
long reach of the spiral
 waves \citepalias{2019arXiv190802326D}.

For comparison, we  show in the top panel of Figure \ref{fig:planet_vss} the result from an 
Eulerian simulation, done with {\footnotesize FARGO3D}.  The setup  is very similar to the simulations in \citetalias{2019arXiv190802326D}.
The grid is logarithmically-spaced in radius, and extends from $r = [0.5, 35] r_0$ with $N_r, N_\phi = (340,502)$. It also
 includes a wave-killing zone inside of $r_0$. 
 As seen in the figure, the agreement between {\footnotesize FARGO3D} and {\footnotesize AREPO} is excellent.

We turn now to examining the torques in the {\footnotesize AREPO} simulation.  One of the key
quantities in studying planet-disk interaction is the value of the net torque $\Delta T$
(Equation \ref{eq:deltat_def}) far beyond the planet's
orbit.
Recall that $\Delta T$ 
is one of the two constants in a steady-state planet-less disk. Therefore far beyond the planet's
orbit $\Delta T$ should be spatially constant in viscous steady state.  The value of that constant determines
both the planet's migration rate  and the ``pileup'' of material outside of its  orbit \citepalias{2019arXiv190802326D}.

To set the stage for the torques in the planet simulation, we first show in
the middle panel of Figure \ref{fig:planet_vss}  what happens {\it without} a planet, 
in one of the simulations from Figure \ref{fig:2d_1d_comp}. We see that $\Delta T$ vanishes
at $r_0$ and inwards, demonstrating that our torque-free mass removal procedure is behaving as
intended.  In addition, $\Delta T$ remains negligibly small out to $r\sim 10r_0$, demonstrating
that the disk is in viscous steady state out to that radius\footnote{
The rise in $\Delta T$ beyond $10r_0$ is caused by the outer exponential cutoff with which 
we initialized the simulation.}.  We also show in the figure the two components
of $\Delta T$, i.e., $F_\nu$ and $\dot{M}\ell$, and these are both equal to $\dot{M}_{\rm acc}\ell\propto r^{1/2}$ throughout the physical part of the domain that is in steady state, as required
by the planet-less viscous steady state solution (\S \ref{sec:why}); 
 here,  $\dot{M}_{\rm acc}$ is the rate at which mass is artificially removed within the inner zone
 of the simulation. 
 
The torques in the {\footnotesize AREPO} planet simulation are shown in the bottom panel  of Figure \ref{fig:planet_vss}, where
 they are time-averaged over  $50$ planetary orbits, and the $1$-$\sigma$ deviations are shown as  shaded regions.  We observe again that $\Delta T$ is negligibly small at
$r_0$, as required, and remains small until near the planet's orbit.
One may note that $\Delta T$ does not reach a constant value at large $r$, because the simulation is
not quite in viscous steady state there (similar to the middle panel).
But given that the constant's value is so important, we display it in a different way:
we consider $T_{\rm dep}(r)$, which is the torque deposited into the disk by spiral  waves that have been launched
by the planet \citepalias{2019arXiv190802326D}. 
In viscous steady state with a planet, Equation (\ref{eq:deltat_def})  is modified to
$\Delta T(r)=T_{\rm dep}(r)$, i.e., 
 the net torque in the disk must equal that provided by the deposition of spiral wave torque. 
The figure shows $T_{\rm dep}(r)$, which we measure
by combining the  torque required to excite the waves with
the  torque transported   by the  waves (as explained in e.g., \citealt{2017PASJ...69...97K}; \citetalias{2019arXiv190802326D}). 

We see that $\Delta T\approx T_{\rm dep}$ within $\sim 10 r_0$, as required by viscous
steady state. 
In addition, 
  the total torque deposited into the disk, i.e., $T_{\rm dep}$ at ${r\gg r_{\rm pl}}$, is 
  very small in absolute value relative to   $\dot{M}_{\rm acc}\ell_0$.  Moreover, 
   further investigation shows  that it oscillates with small amplitude over time, likely
  due to the disk being slightly eccentric (Dempsey et al., in prep). We  conclude
  that for this setup the planet migration rate is very slow, and moreover there is a negligibly
  small pileup outside of the planet's orbit.

\section{Summary and Discussion} \label{sec:conclusions}

Our main results are as follows:
\begin{itemize}
	\item The inner boundaries of accretion disks should be designed to ensure that they are
	torque free ($\Delta T=0$). Otherwise, the density profile beyond the boundary will be incorrect.
	Conversely,  one may infer that 
	   boundary-induced density depressions found in many quasi-Lagrangian simulations
	 are due to
	  an artificial torquing of the disk by the mass removal algorithm. 
		
	\item We introduced a simple method to ensure a torque-free inner boundary in quasi-Lagrangian
	codes that use mass removal schemes:
	 the azimuthal velocity of the  material remaining after mass removal is increased to avoid changing
	 the angular momentum. 
	Once the angular momentum is properly handled, further details of mass removal have 
	 negligible impact on the disk's behavior. 
	
	\item We implemented this mass removal in {\footnotesize AREPO}, showing that it produced the
	desired behavior both in planet-free disks, and in a disk with a planet---allowing one
	to determine the total torque of the planet on the disk, $\Delta T$.

\end{itemize}

Our mass removal algorithm is similar  to the angular 
momentum-preserving sinks of \citet{2004ApJ...611..399K}. 
But the  disks in  \citet{2004ApJ...611..399K} do not achieve the ZAM solution
outside the accretion zone (cf. their Figures 7-8). 
Instead, they exhibit an ``evacuation zone'' that grows in time, implying
that their algorithm is not truly torque-free.  We hypothesize that this might be caused by inadequate
resolution.

\citet{1995MNRAS.277..362B} introduced a fix to correct for the discontinuity in density
across the inner boundary present in SPH simulations. Their fix appears to alleviate at least part of the boundary-induced density depletion (see their Figure 2). However, it does not ensure that $\Delta T=0$ in the inner parts of the simulation.

Although we have focused on disks nearly in viscous steady state, inner boundaries should
almost always be torque-free, even for time-dependent disks.  That is because the viscous timescale
is shortest in the inner regions, and so even if the disk is evolving viscously, the innermost regions
should be effectively in steady state.  The only exception we see 
is when one wishes to model the effect of a true inner boundary---e.g., due to a stellar magnetosphere. 
Nonetheless, we do not claim that previous simulations with non-torque-free inner boundaries
are  invalid.  Far away from such a boundary, the boundary's effect may indeed be
ignored.

We foresee a number of applications.  The planet
simulation considered here may be extended to other cases, including binary stars and more realistic disks. 
For that simulation, {\footnotesize FARGO3D} was sufficient---indeed, it is around five times as fast as {\footnotesize AREPO}\footnote{ For this comparison {\footnotesize AREPO} was run on 28 CPU cores with $\sim 140,000$ zones and {\footnotesize FARGO3D} was run on one NVIDIA V100 GPU with $\sim 170,000$ zones.} . 
But for cases where the flow becomes strongly non-Keplerian, the orbital advection algorithm used in {\footnotesize FARGO3D} becomes much less efficient. 
Additional applications include modeling of the circumplanetary/circumsecondary  disk region
  and the evolution of warps. 
Regarding the last application, we foresee no great difficulty in extending our 2D method to 3D ---although the requirements for our method to work in 3D have yet to be examined.

\acknowledgements 
We would like to thank the anonymous referee for their helpful comments.
YL acknowledges NASA grant NNX14AD21G and NSF grant AST1352369. 
This work used computing resources provided by Northwestern University and the Center for Interdisciplinary Exploration and Research in Astrophysics (CIERA) funded by NSF PHY-1726951, and was supported in part through the computational resources and staff contributions provided for the Quest high performance computing facility at Northwestern University which is jointly supported by the Office of the Provost, the Office for Research, and Northwestern University Information Technology.

\appendix

\section{{\footnotesize AREPO} Simulation Details} \label{sec:app_2d}

\subsection{Refinement Criteria} \label{sec:app_ref}

To ensure that there is adequate spatial resolution, we make use of two different cell refinement/derefinement criteria. 
The first, which becomes most important in the vicinity of the planet, is the default mass-based resolution in {\footnotesize AREPO}.  We specify a target cell mass, $m_{\rm target}$.
If a cell has a mass outside of the range $m_{\rm cell} = [0.5 m_{\rm target}, 2 m_{\rm target}]$, that cell is either split or merged with surrounding cells so that the resulting mass falls within the specified range. 
In our  simulation with a planet, we smoothly decrease $m_{\rm target}$ by a factor of 300 in the Hill sphere of the planet.  

Our second refinement criteria, which becomes important in the vicinity of the star, and is used in conjunction with the mass based criterion, is cell size based and  enforces a minimum number of cells in the azimuthal direction.
We specify a target cell radius, $r_{\rm target}$, that is equal to $2 \pi r / N_\phi$, where $N_\phi$ is the desired number of cells in the azimuthal direction. 
If a cell has a radius outside of the range $r_{\rm cell} =[0.5 r_{\rm target}, 2 r_{\rm cell}]$, it is either split or joined with neighboring cells until the cell radius falls within our specified range. 
To ensure a smooth transition in cell size towards the star we lower the $N_\phi$ from the nominal target value as $(r/r_0)^3$ inside $r_0$.

\subsection{Resolution Requirements} \label{sec:app_res}

\begin{figure*}
    \centering
    \includegraphics[trim={0.2cm .2cm 0.2cm 0},clip,width=.98\textwidth]{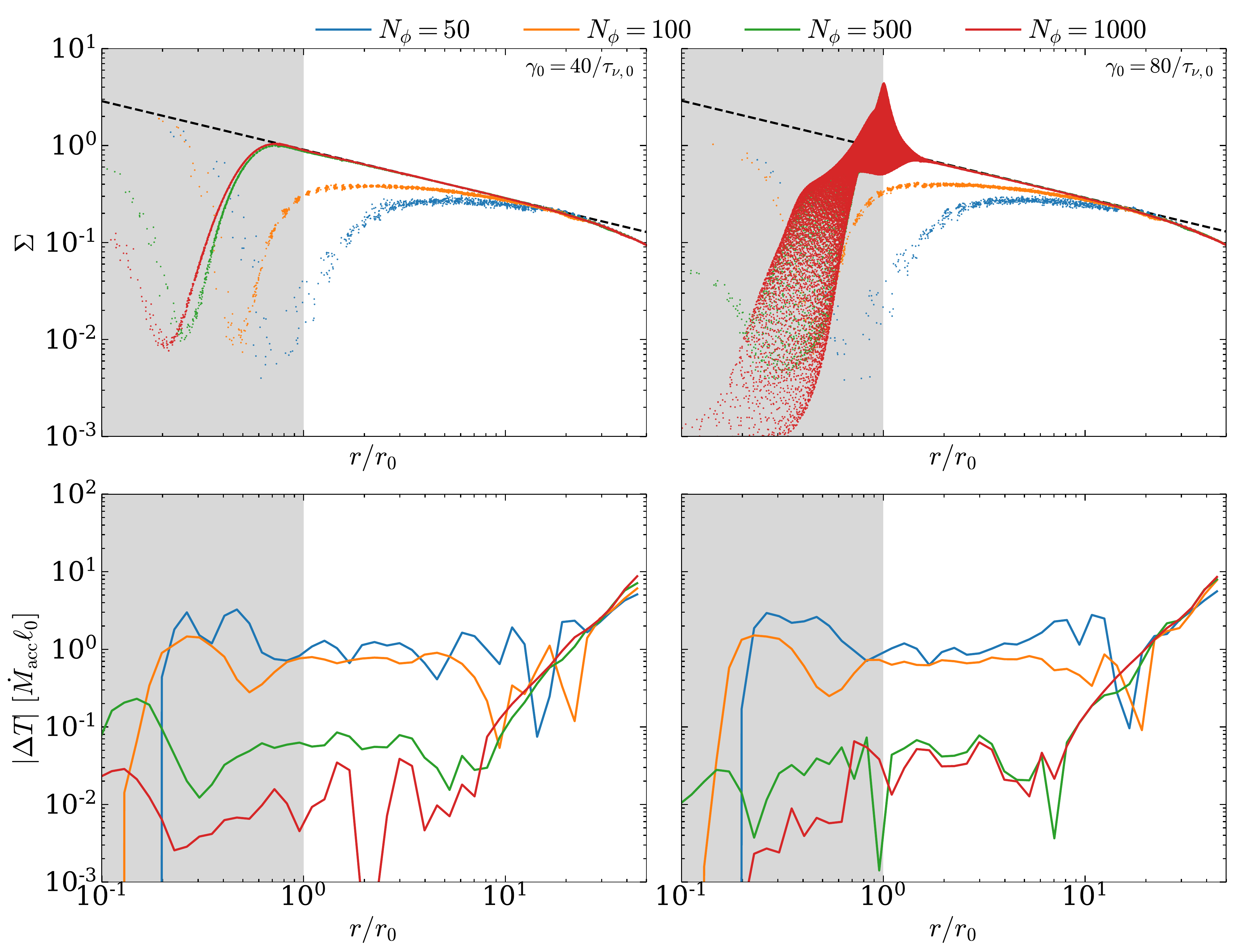}
    \caption{The resolution requirements for torque-free mass removal in planet-less 
    {\footnotesize AREPO} simulations.
    The two left panels show the case with $\gamma_0=40/\tau_{\nu,0}$.
     In the top-left panel, the dots show  $\Sigma$ within every cell, in which case non-axisymmetries
      are evident as the spread in points at a given $r$. 
    We see that one requires $N_\phi\gtrsim $a few hundred
    for the steady state profile to match the ZAM solution (dashed line). 
    The lower-left panel shows the net torque. We see that even 
    for the crudest resolution $\Delta T$ is nearly spatially constant. The error in the top-left
    panel is therefore due to the errors within the mass removal zone (which can likely
    be corrected with a better treatment of that zone).  
    The right panels show a case with $\gamma_0$ twice as large. In this case, 
    the disk becomes highly non-axisymmetric at a resolution $N_\phi=500$; surprisingly, however, 
    that does not much affect the behavior further out. 
    } 
    \label{fig:arepo_resolution}
\end{figure*}

We explore different values of $N_\phi$ in Figure \ref{fig:arepo_resolution} for planet-less disks, with two different $\gamma_0$ values. 
We conclude that $\gamma_0=40/\tau_{\nu,0}$ and $N_\phi=1000$ (the values used in the
body of this {\it Letter}) suffice to accurately simulate the disk.

\subsection{Flux measurements} \label{sec:fluxes}

In Figure \ref{fig:planet_vss} we show the azimuthally-averaged steady-state torques. 
Computing these azimuthal averages is a non-trivial task in {\footnotesize AREPO} as the cells are not on a cylindrical grid. 
During a simulation, we add a cell's content (be it mass, viscous flux, $\dot{M}$, etc.) into finely spaced radial bins. 
For each binned quantity, we compute the cumulative radial integral across the domain, and then compute the numerical derivative on a coarser grid. 
This leaves us with an estimate for the azimuthal average of each quantity. 
For the planet-less simulations we use logarithmically spaced bins with width $\Delta \ln r_{\rm bin} \approx 0.001$ and evaluate the numerical derivatives with a radial grid with spacing $\approx 0.14$. 
For the planet simulation we evaluate the numerical derivatives with a finer radial grid with spacing $\approx 0.034$. 


\begin{thebibliography}{}
\expandafter\ifx\csname natexlab\endcsname\relax\def\natexlab#1{#1}\fi
\providecommand{\url}[1]{\href{#1}{#1}}
\providecommand{\dodoi}[1]{doi:~\href{http://doi.org/#1}{\nolinkurl{#1}}}
\providecommand{\doeprint}[1]{\href{http://ascl.net/#1}{\nolinkurl{http://ascl.net/#1}}}
\providecommand{\doarXiv}[1]{\href{https://arxiv.org/abs/#1}{\nolinkurl{https://arxiv.org/abs/#1}}}

\bibitem[{Bate {et~al.}(1995)Bate, Bonnell, \& Price}]{1995MNRAS.277..362B}
Bate, M.~R., Bonnell, I.~A., \& Price, N.~M. 1995, \mnras, 277, 362


\bibitem[{Ben{\'\i}tez-Llambay \& Masset(2016)}]{2016ApJS..223...11B}
Ben{\'\i}tez-Llambay, P., \& Masset, F.~S. 2016, \apjs, 223, 11

\bibitem[{Chandrasekhar(1961)}]{1961hhs..book.....C}
Chandrasekhar, S. 1961, International Series of Monographs on Physics

\bibitem[{de~Val-Borro {et~al.}(2006)de~Val-Borro, Edgar, Artymowicz,
  Ciecielag, Cresswell, D'Angelo, Delgado-Donate, Dirksen, Fromang,
  Gawryszczak, Klahr, Kley, Lyra, Masset, Mellema, Nelson, Paardekooper,
  Peplinski, Pierens, Plewa, Rice, Sch{\"a}fer, \&
  Speith}]{2006MNRAS.370..529D}
de~Val-Borro, M., Edgar, R.~G., Artymowicz, P., {et~al.} 2006, \mnras, 370, 529


\bibitem[{Dempsey {et~al.}(2020)Dempsey, Lee, \&
  Lithwick}]{2019arXiv190802326D}
Dempsey, A.~M., Lee, W.-K., \& Lithwick, Y., 2020, \apj, 891, 108


\bibitem[{Duffell {et~al.}(2019)Duffell, D'Orazio, Derdzinski, Haiman,
  MacFadyen, Rosen, \& Zrake}]{Duffell:2019vu}
Duffell, P.~C., D'Orazio, D., Derdzinski, A., {et~al.} 2019, arXiv,
  arXiv:1911.05506

\bibitem[{Farris {et~al.}(2014)Farris, Duffell, MacFadyen, \&
  Haiman}]{2014ApJ...783..134F}
Farris, B.~D., Duffell, P., MacFadyen, A.~I., \& Haiman, Z. 2014, \apj, 783,
  134

\bibitem[{Goldreich \& Tremaine(1980)}]{1980ApJ...241..425G}
Goldreich, P., \& Tremaine, S. 1980, \apj, 241, 425

\bibitem[{Hopkins(2017)}]{2017arXiv171201294H}
Hopkins, P.~F. 2017, arXiv, arXiv:1712.01294

\bibitem[{Hubber {et~al.}(2018)Hubber, Rosotti, \& Booth}]{2018MNRAS.473.1603H}
Hubber, D.~A., Rosotti, G.~P., \& Booth, R.~A. 2018, \mnras, 473, 1603

\bibitem[{Kanagawa {et~al.}(2017)Kanagawa, Tanaka, Muto, \&
  Tanigawa}]{2017PASJ...69...97K}
Kanagawa, K.~D., Tanaka, H., Muto, T., \& Tanigawa, T. 2017, \pasj, 69, 97

\bibitem[{Krumholz {et~al.}(2004)Krumholz, McKee, \&
  Klein}]{2004ApJ...611..399K}
Krumholz, M.~R., McKee, C.~F., \& Klein, R.~I. 2004, \apj, 611, 399

\bibitem[{Li {et~al.}(2000)Li, Finn, Lovelace, \&
  Colgate}]{2000ApJ...533.1023L}
Li, H., Finn, J.~M., Lovelace, R. V.~E., \& Colgate, S.~A. 2000, \apj, 533,
  1023

\bibitem[{Lovelace {et~al.}(1999)Lovelace, Li, Colgate, \&
  Nelson}]{1999ApJ...513..805L}
Lovelace, R. V.~E., Li, H., Colgate, S.~A., \& Nelson, A.~F. 1999, \apj, 513,
  805

\bibitem[{Lunine \& Stevenson(1982)}]{1982Icar...52...14L}
Lunine, J.~I., \& Stevenson, D.~J. 1982, \icarus, 52, 14

\bibitem[{Lynden-Bell \& Pringle(1974)}]{1974MNRAS.168..603L}
Lynden-Bell, D., \& Pringle, J.~E. 1974, \mnras, 168, 603

\bibitem[{Moody {et~al.}(2019)Moody, Shi, \& Stone}]{2019ApJ...875...66M}
Moody, M. S.~L., Shi, J.-M., \& Stone, J.~M. 2019, \apj, 875, 66

\bibitem[{Mu{\~n}oz {et~al.}(2014)Mu{\~n}oz, Kratter, Springel, \&
  Hernquist}]{2014MNRAS.445.3475M}
Mu{\~n}oz, D.~J., Kratter, K., Springel, V., \& Hernquist, L. 2014, \mnras,
  445, 3475

\bibitem[{Mu{\~n}oz \& Lai(2016)}]{2016ApJ...827...43M}
Mu{\~n}oz, D.~J., \& Lai, D. 2016, \apj, 827, 43

\bibitem[{Mu{\~n}oz {et~al.}(2020)Mu{\~n}oz, Lai, Kratter, \&
  Miranda}]{2020ApJ...889..114M}
Mu{\~n}oz, D.~J., Lai, D., Kratter, K., \& Miranda, R. 2020, \apj, 889, 114

\bibitem[{Mu{\~n}oz {et~al.}(2019)Mu{\~n}oz, Miranda, \&
  Lai}]{2019ApJ...871...84M}
Mu{\~n}oz, D.~J., Miranda, R., \& Lai, D. 2019, \apj, 871, 84

\bibitem[{Mu{\~n}oz {et~al.}(2013)Mu{\~n}oz, Springel, Marcus, Vogelsberger, \&
  Hernquist}]{2013MNRAS.428..254M}
Mu{\~n}oz, D.~J., Springel, V., Marcus, R., Vogelsberger, M., \& Hernquist, L.
  2013, \mnras, 428, 254

\bibitem[{Pakmor {et~al.}(2016)Pakmor, Springel, Bauer, Mocz, Mu{\~n}oz,
  Ohlmann, Schaal, \& Zhu}]{2016MNRAS.455.1134P}
Pakmor, R., Springel, V., Bauer, A., {et~al.} 2016, \mnras, 455, 1134

\bibitem[{Price {et~al.}(2018)Price, Wurster, Tricco, Nixon, Toupin, Pettitt,
  Chan, Mentiplay, Laibe, Glover, Dobbs, Nealon, Liptai, Worpel, Bonnerot,
  Dipierro, Ballabio, Ragusa, Federrath, Iaconi, Reichardt, Forgan, Hutchison,
  Constantino, Ayliffe, Hirsh, \& Lodato}]{2018PASA...35...31P}
Price, D.~J., Wurster, J., Tricco, T.~S., {et~al.} 2018, \pasa, 35, e031

\bibitem[{Shakura \& Sunyaev(1973)}]{1973A&A....24..337S}
Shakura, N.~I., \& Sunyaev, R.~A. 1973, \aap, 24, 337

\bibitem[{Springel(2010)}]{2010MNRAS.401..791S}
Springel, V. 2010, \mnras, 401, 791

\bibitem[{Springel {et~al.}(2001)Springel, Yoshida, \&
  White}]{2001NewA....6...79S}
Springel, V., Yoshida, N., \& White, S. D.~M. 2001, New Astronomy, 6, 79

\bibitem[{Tang {et~al.}(2017)Tang, MacFadyen, \& Haiman}]{2017MNRAS.469.4258T}
Tang, Y., MacFadyen, A., \& Haiman, Z. 2017, \mnras, 469, 4258

\end{thebibliography}
\end{document}